\documentclass{PoS}

\newcommand{\beq}{\begin{eqnarray}}
\newcommand{\eeq}{\end{eqnarray}}

\newcommand{\Lams}{\Lambda_{\overline{\rm MS}}}

\newcommand{\be}{\begin{equation}}
\newcommand{\ee}{\end{equation}}
\newcommand{\lwrsim}{\raise0.3ex\hbox{$<$\kern-0.75em\raise-1.1ex\hbox{$\sim$}}}
\newcommand{\lgrsim}{\raise0.3ex\hbox{$>$\kern-0.75em\raise-1.1ex\hbox{$\sim$}}}

\def\C2#1#2{({\cal C}_2)_{#1}^{#2}}

\def\eq#1{eq.~(\ref{#1})}

\def\VEV#1{\langle #1 \rangle}

\title{Ghost-gluon coupling, power corrections and $\Lambda_{\overline{MS}}$
from twisted-mass lattice QCD at $N_f=2$}

\ShortTitle{Ghost-gluon coupling, power corrections and $\Lambda_{\overline{MS}}$
from twisted-mass lattice QCD at $N_f=2$}

\author{B. Blossier, Ph. Boucaud, \speaker{M. Gravina}, O. P\`ene\\
        Laboratoire de Physique Th\'eorique\footnote{Unit\'e Mixte 
        de Recherche 8627 du Centre National de
        la Recherche Scientifique}\\
        {Universit\'e de Paris-Sud, B\^atiment 210, 91405 Orsay Cedex,
        France}\\
        E-mail: \email{Benoit.Blossier,Philippe.Boucaud,Mario.Gravina,Olivier.Pene@th.u-psud.fr}}

\author{F. De soto\\
        Dpto. Sistemas F\'isicos, Qu\'imicos y Naturales, \\ 
        Universidad Pablo de Olavide, 41013 Sevilla, Spain.\\
        E-mail: \email{fcsotbor@upo.es}}

\author{V. Morenas\\
        Laboratoire de Physique Corpusculaire,\\ 
        Universit\'e Blaise Pascal, CNRS/IN2P3, France\\
        E-mail: \email{morenas@in2p3.fr}}

\author{J. Rodr\'iguez-Quintero\\
        Dpto. F\'isica Aplicada, Fac. Ciencias Experimentales,\\
        Universidad de Huelva, 21071 Huelva, Spain\\
        E-mail: \email{jose.rodriguez@dfaie.uhu.es}}

\abstract{
\begin{center}
\raisebox{5mm}{\hbox{
\includegraphics[width=3cm]{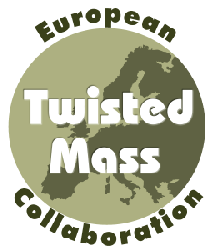}}}
\end{center}

A non-perturbative calculation of the ghost-gluon running QCD coupling
constant is performed using $N_f=2$ twisted-mass dynamical fermions.
The extraction of $\Lambda_{\overline{MS}}$ in the chiral limit 
reveals the presence of a non-perturbative OPE contribution that is 
assumed to be dominated by a dimension-two $\VEV{A^2}$ condensate.
In this contest a novel method for calibrating the lattice spacing 
in lattice simulations is presented.

}

\FullConference{The XXVIII International Symposium on Lattice Field Theory, Lattice2010\\
                June 14-19, 2010\\
                Villasimius, Italy}

\begin{document}

\section{Introduction}

$\Lambda_{\overline{MS}}$ is the scale of strong interactions. This parameter has to be 
taken from experiment and can be determined from the running of the QCD coupling constant.
This latter can be calculated in a variety of non-perturbative ways on the lattice (see ~\cite{Luscher:1993gh,Gockeler:2005rv,Alles:1996ka,Boucaud:tanti,Sternbeck:2007br} and references therein).
In the {\it quenched} case~\cite{Boucaud:2008gn} the comparison between the perturbative and lattice determinations
over a large momentum window revealed the presence of a dimension-two $\VEV{A^2}$ condensate,
signaling that momenta considered in lattice simulation are in a non-perturbative region. 
Here we extend the strategy of \cite{Boucaud:2008xu} to the case of $N_f=2$ twisted mass in the sea sector
using configurations produced by the ETM Collaboration~\cite{Boucaud:2008xu}, in order to study the effect of 
the quark mass.

\section{Lattice computation of the coupling in the Taylor scheme}

Following~\cite{Boucaud:2008gn}, we calculate the strong coupling constant from the ghost-gluon vertex. 
Gluon and ghost propagartors in the Landau gauge are defined as
\beq\label{props}
\left( G^{(2)} \right)_{\mu \nu}^{a b}(p^2,\Lambda) = \frac{G(p^2,\Lambda)}{p^2} \ \delta_{a b}
\left( \delta_{\mu \nu}-\frac{p_\mu p_\nu}{p^2} \right) \ ,
\left(F^{(2)} \right)^{a,b}(p^2,\Lambda) = - \delta_{a b} \ \frac{F(p^2,\Lambda)}{p^2} \ 
\eeq
where $\Lambda=a^{-1}(\beta)$ is the regularisation cut-off. $G$ and $F$ are the gluon and
ghost dressing functions which can be determined by a non-perturbative renormalization (MOM).
In the Taylor scheme~\cite{Taylor}, where the incoming ghost momentum vanishes, the ghost-gluon vertex
does not renormalize. This allows for a simple determination of the renormalized coupling constant 
in this scheme as
\beq\label{alpha}
\alpha_T(\mu^2) \equiv \frac{g^2_T(\mu^2)}{4 \pi}=  \ \lim_{\Lambda \to \infty}
\frac{g_0^2(\Lambda^2)}{4 \pi} G(\mu^2,\Lambda^2) F^{2}(\mu^2,\Lambda^2) \ ;
\eeq
in terms of only two-point gluon and ghost dressing function. Here $g_0$ is the bare strong
coupling and $\mu$ the renormalization scale.
This definition can be used in a lattice determination and is to be compared with a theoretical 
formula in order to extract $\Lambda_{\rm QCD}$. As in the {\it quenched} case, using the four-loops 
expression for the coupling constant in the Tayol scheme~\cite{Chetyrkin00,Chetyrkin:2004mf}
\beq
  \label{betainvert}
      \alpha_T(\mu^2) &=& \frac{4 \pi}{\beta_{0}t}
      \left(1 - \frac{\beta_{1}}{\beta_{0}^{2}}\frac{\log(t)}{t}
     + \frac{\beta_{1}^{2}}{\beta_{0}^{4}}
       \frac{1}{t^{2}}\left(\left(\log(t)-\frac{1}{2}\right)^{2}
     + \frac{\widetilde{\beta}_{2}\beta_{0}}{\beta_{1}^{2}}-\frac{5}{4}\right)\right) \nonumber \\
     &+& \frac{1}{(\beta_{0}t)^{4}}
 \left(\frac{\widetilde{\beta}_{3}}{2\beta_{0}}+
   \frac{1}{2}\left(\frac{\beta_{1}}{\beta_{0}}\right)^{3}
   \left(-2\log^{3}(t)+5\log^{2}(t)+
\left(4-6\frac{\widetilde{\beta}_{2}\beta_{0}}{\beta_{1}^{2}}\right)\log(t)-1\right)\right)
\eeq
where $t=\ln \frac{\mu^2}{\Lambda_T^2}$ and coefficients are
\beq\label{betacoefs}
\beta_0 &=&  11 - \frac 2 3 N_f \ , \ \
\beta_1 = \overline{\beta}_1 = 102 - \frac{38} 3 N_f
\nonumber \\
\widetilde{\beta}_2 &=& 3040.48 \ - \ 625.387 \ N_f \ + \ 19.3833 \ N_f^2
\nonumber \\
\widetilde{\beta}_3 &=& 100541 \ - \ 24423.3 \ N_f \ + \ 1625.4 \ N_f^2 \ - \ 27.493 \ N_f^3
\ ,
\eeq
Extracting $\Lambda_T$ from the lattice data at each $\mu^2$ using this perturbative
formula does not lead to a constant value.
To understand the mismatch beetween lattice and perturbative determination, a non-perturbative OPE 
correction to the perturbative formula is to be considered. This accounts for the minimal power 
correction associated to the presence of a dimension-two $\VEV{A^2}$ condensate:
\beq\label{alphahNPt}
\alpha_T(\mu^2) =
\alpha^{\rm pert}_T(\mu^2)
\ \left(  1 + \frac{9}{\mu^2} \frac{g^2_T(q_0^2) \langle A^2 \rangle_{R,q_0^2}} {4 (N_C^2-1)}
\right) \ ,
\eeq
where $q_0^2 \gg \Lambda_{\rm QCD}$ is some perturbative scale.
This will cure the mismatch and lead to a good determination for $\Lambda_{T}$
in the Taylor scheme, which eventually can be be related to the value of the scale
in the $\overline{MS}$ scheme through
\beq\label{ratTMS}
\frac{\Lambda_{\overline{\rm MS}}}{\Lambda_T} \ = \ e^{\displaystyle -\frac{c_1}{2 \beta_0}} \ = \
e^{\displaystyle - \frac{507-40 N_f}{792 - 48 N_f}}\ = \ 0.541449
\ .
\eeq

\section{Artefacts}

We exploited data from ETMC configurations at maximal twist for a variety of run 
parameters (tab.~\ref{tables}) 
in order to study physical and systematic effects in our determinations.
This have the main advantage of reducing the discretization artefacts to $\mathcal{O}(a^2)$, 
where $a$ is the lattice spacing.
Nevertheless, artefacts are expected to came at different levels.
A first kind of artefacts that can be systematically
cured~\cite{Becirevic:1999uc,deSoto:2007ht} are those due to the
breaking of the  rotational symmetry of the euclidean space-time when using an
hypercubic lattice,  where this symmetry is restricted to the discrete $H(4)$
isometry group. It is convenient to compute first the average  of any
dimensionless lattice quantity $Q(a p_\mu)$ over every orbit of the group
$H(4)$. In general several orbits of $H(4)$ correspond to one value of $p^2$.
Defining the $H(4)$ invariants $p^{[n]}=\sum_{\mu=1}^{4} p_\mu^n$, if the 
lattice spacing is small enough such that $\epsilon=a^2 p^{[4]}/p^2<<1$,
the dimensionless lattice correlation function can be
expanded in powers of $\epsilon$:
\beq
Q(a^2\,{p}^2, a^4p^{[4]}, a^6 p^{[6]},a^2\Lambda_{\rm QCD}^2)
= Q(a^2p^2,a^2\Lambda_{\rm QCD}^2) +
\left.\frac{dQ}{d\epsilon}\right|_{\epsilon=0} a^2
\frac{p^{[4]}}{p^2} + \cdots
\eeq
$H(4)$ methods are based on the appearance of a $\mathcal{O}(a^2)$
corrections driven by a $p^{[4]}$ term. The basic method is to fit
from the whole set of orbits sharing the same $p^2$ the
coefficient $dQ/d\epsilon$ and get the extrapolated value of $Q$,
free from $H(4)$ artefacts.

\begin{figure}[hbt]
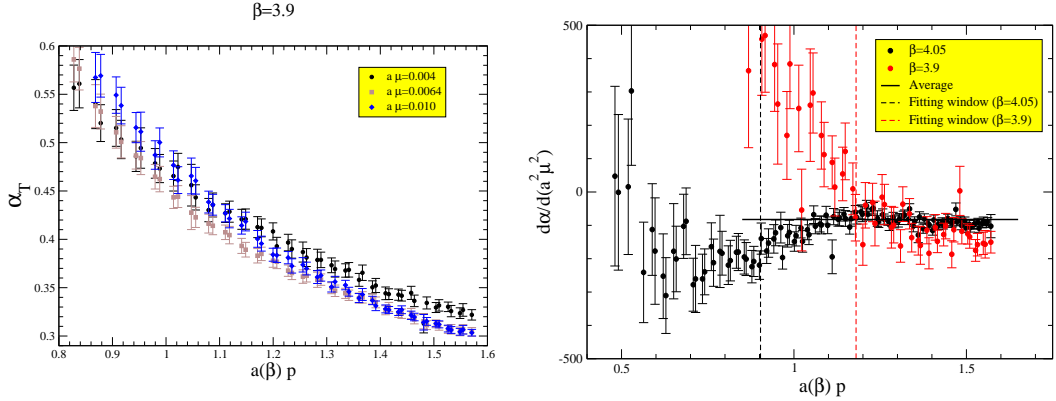

\begin{center}
\begin{tabular}{cc}
\includegraphics[width=6.5cm]{aMOMb39.eps}
&
\includegraphics[width=7cm,height=5.5cm]{C0.eps}
\end{tabular}
\end{center}
\caption{\small Left: The Taylor couplings estimates, after $H(4)$-extrapolation,
at $\beta=3.9$ for $\mu_q=0.004,0.0064,0.010$. Right: The slopes for the mass squared extrapolation in terms of $a p$
computed for the four lattices simulations at $\beta=4.05$ ($32^3\times 64$) with $a \mu_q=0.\
003,0.006,0.008,0.012$ and
for the three ones at $\beta=3.9$ ($24^3\times 48$) with $a \mu_q=0.004,0.0064,0.010$.
}
\label{fig:brut}
\end{figure}

\begin{figure}[hbt]
\begin{center}
\begin{tabular}{ccc}
\includegraphics[width=5.25cm]{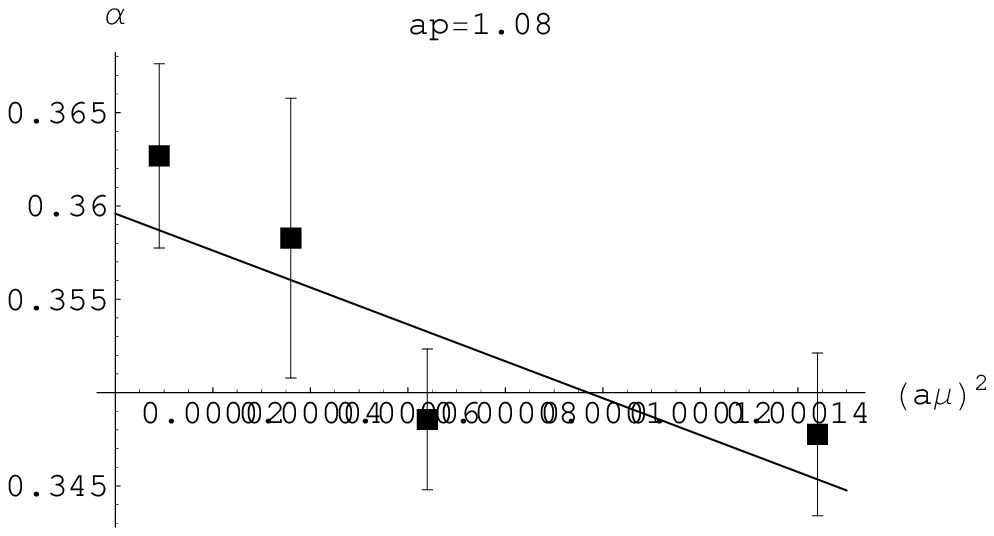}
&
\includegraphics[width=5.25cm]{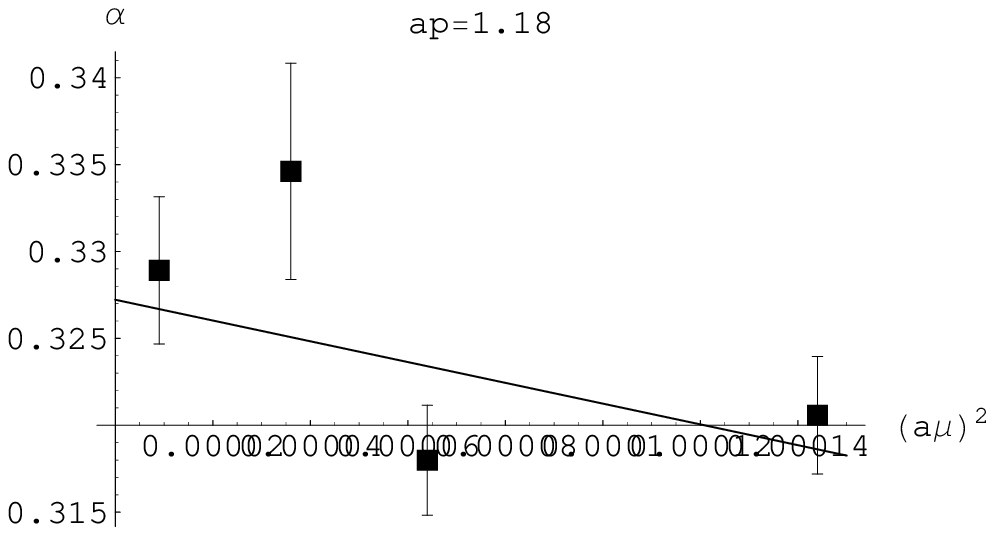}
&
\includegraphics[width=5.25cm]{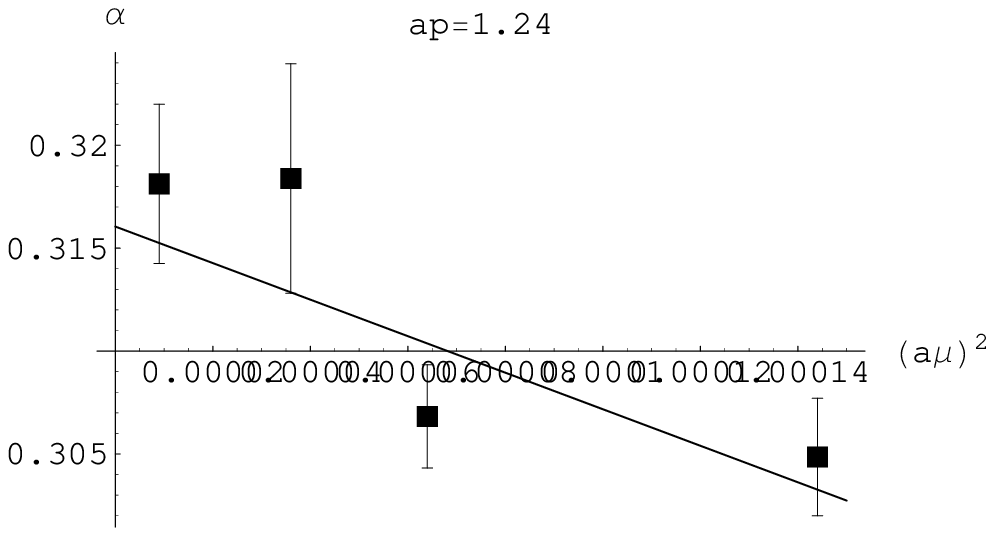}

\end{tabular}
\end{center}
\caption{\small We plot the values of the Taylor coupling at $\beta=4.05$, computed
for some representative values of the lattice momentum, $a(4.05) p=1.08,1.18,1.24,1.36,1.45,1.52$,
in terms of $a^2(4.05) \mu_q^2$ and show the suggested linear extrapolation at $a^2\mu_q^2=0$.}
\label{fig:slopes}
\end{figure}

\begin{figure}[hbt]
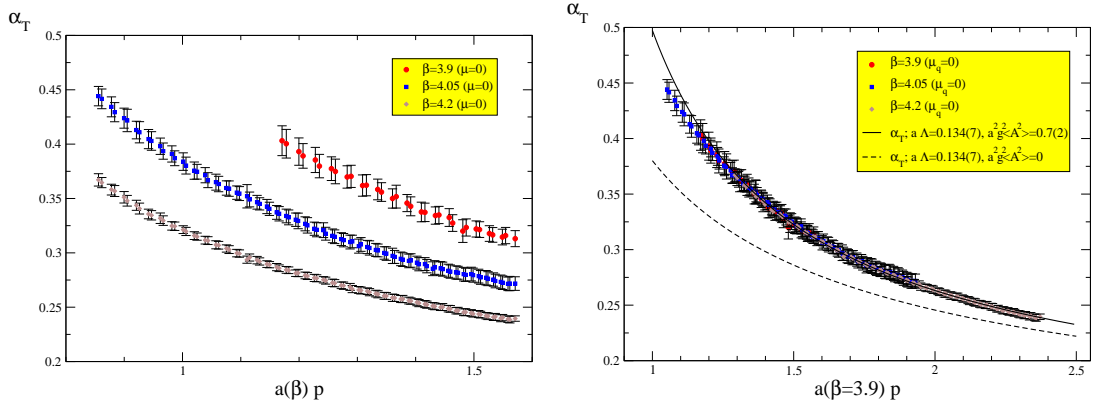

\begin{center}

\begin{tabular}{cc}
\includegraphics[width=7cm]{alpha.eps}
&
\includegraphics[width=7cm,height=5.3cm]{alphaR.eps}
\end{tabular}
\end{center}
\caption{\small Left: The Taylor coupling, free of $H(4)$ and
mass-quarks artefacts, for the three
$\beta=3.9,4.05,4.2$ and plotted in terms of  the lattice momentum $a(\beta) p$.
Right: The scaling of the Taylor coupling computed by
for the three $\beta=3.9,4.05,4.2$ is shown. The lattice momentum,
$a(\beta)p$ in the x-axis, is converted to a physical momentum in units
(the same for the three $\beta$'s) of $a(3.9)^{-1}$.
}
\label{fig:C0}
\end{figure}

A second kinf of artefact could come from dynamical quark masses. 
We will argue that this is a ${\mathcal O(a^2 \mu_q^2)}$ effect and therefore
that it is a lattice artefact. We have calculated
the $H(4)$-free ghost and gluon dressing functions and combined in order to calculate
the $H(4)$-free lattice coupling through~\eq{alpha}. 
In Fig.~\ref{fig:brut} one can see the Taylor coupling after  hypercubic
extrapolation for different $\mu_q$ at fixed $\beta=3.9$ and $4.05$. Indeed,
a dependence in $\mu_q$ is clearly seen. If it is an artefact the dependence
should be in $a^2 \mu_q^2$. If it is an effect in the continuum it should be
some unknown function of the physical mass $\mu_q$.
Trying an $\mathcal{O}(a^2 \mu_q^2)$ dependence, we write the expansion :
\beq\label{H4exp}
\widehat{\alpha}_T(a^2p^2,a^2\mu_q^2)
= \alpha_T(p^2) + R_0(a^2p^2) \ a^2 \mu_q^2 , \ \
R_0(a^2 p^2) \equiv
\frac{\partial \widehat{\alpha}_{T}}{\partial (a^2\mu_q^2)} \
\eeq
Provided that the first-order expansion in \eq{H4exp} is reliable,  a linear
behaviour on $a^2 \mu_q^2$ has to be expected for the  lattice estimates of
$\widehat{\alpha}_T$ for any fixed lattice  momentum computed from simulations
at any given $\beta$ and several   values of $\mu_q$. We explicitely check this
linear behaviour to occur  for the results from our $\beta=4.05$ and $\beta=3.9$
simulations and  show in Fig.~\ref{fig:slopes} some plots of
$\widehat{\alpha}_T$ computed at $\beta=4.05$  (where four different quark masses
are available) for some representatives lattice momenta  in terms of
$a^2\mu_q^2$.
In fig.~\ref{fig:C0}, we plot $R_0(a^2p^2)$ as a function of
 $a p$ computed for the four lattices simulations at $\beta=4.05$ with
different quark masses and  for the three ones at $\beta=3.9$.  
Indeed, it can be seen that a constant behaviour appears to
be achieved for $ p \ge p_{\rm min} \simeq 2.8$ GeV. We will not risk an
interpretation of the data below $(a p)_{\rm min}$. The striking observation
here is that above $p_{\rm min}$ both lattice spacings exhibit a fairly constant
 $R_0(a^2p^2)$ and a good enough scaling between both $\beta$'s.
 The fact that $R_0$ with our present data  goes to the same constant  for both
$\beta$'s, leads us to consider that the $\mu$ dependence of $\alpha$ is mainly
  a lattice artefact (else it should be a
function of $\mu$ and not of $a\mu$).

The main result of this work is taking into account the effects due to dynamical 
quarks in a global analysis of the lattice determinations. This lead to a proper 
extrapolation to the continuum limit, which can be compared with continuous formula
in order to extract $\Lambda_{\overline{ \rm MS}}$.

\begin{figure}[hbt!]
\begin{center}
\begin{tabular}{cc}
%
\begin{tabular}{||c|c|c|c||}
\hline
\hline
$\beta$ & $a \mu_q$ & Volume & Number of confs.
\\ \hline
$3.9$ &
\begin{tabular}{c}
0.004 \\
0.0064 \\
0.010
\end{tabular}
&
$24^3\times48$ &
\begin{tabular}{c}
$120$ \\
$20$ \\
$20$
\end{tabular}
\\ \hline
$4.05$ &
\begin{tabular}{c}
0.003 \\
0.006 \\
0.008 \\
0.012
\end{tabular}
& $32^3\times64$ &
\begin{tabular}{c}
$20$ \\
$20$ \\
$20$ \\
$20$
\end{tabular}
\\ \hline
$4.2$ & 0.0065 & $32^3\times64$ & $20$
\\ \hline
\hline
\end{tabular}
&
\begin{tabular}{|c|c|c|}
\hline
 & This paper & String tension \\
\hline
$a(3.9)/a(4.05)$ & 1.224(23) & 1.255(42)  \\
\hline
$a(3.9)/a(4.2)$ & 1.510(32)& 1.558(52)  \\
\hline
$a(4.05)/a(4.2)$ & 1.233(25)& 1.241(39) \\
\hline
$\Lams a(3.9)$ & 0.134(7) &  \\
\hline
$g^2 \langle A^2 \rangle a^2(3.9)$ & 0.70(23) & \\
\hline
\end{tabular}
\end{tabular}
\end{center}
\caption{\small Left: Run parameters of the exploited data from ETMC collaboration.
Right: Best-fit parameters for the ratios of lattice spacings, $\Lams$
and the gluon condensate (for which $a(3.9) q_0=4.5$ is chosen). For the sake
of comparison, we also quote the results from \cite{Baron:2009wt} that
were obtained by computing the hadronic quantity, $r_0/a(\beta)$, and applying
to it a chiral extrapolation.
}
\label{tables}
\end{figure}

\section{$\Lambda_{\overline{ \rm MS}}$  and the gluon condensate}

The running of $\alpha_T$ given by the combination of Green
functions in eq.~(\ref{alpha}) and the extrapolation through \eq{H4exp}, provided that
we are not far from the continuum limit and discretization errors are treated properly,
depend only on the momentum (except, maybe, finite volume errors at low momenta).
The supposed scaling of the Taylor coupling implies for the three curves plotted in
fig.~\ref{fig:C0} to match to each other after
the appropriate conversion of the momentum (in x-axis) from lattice to
physical units, with the multiplication by the lattice spacing at each $\beta$.
Thus, we can apply the ``plateau''-method described in~\cite{Boucaud:2008gn}
for the three $\beta$'s all at once by requiring the minimisation of the total $\chi^2$:
\beq\label{chi2T}
\chi^2\left(a(\beta_0)\Lams,c,\frac{a(\beta_1)}{a(\beta_0)},\frac{a(\beta_2)}{a(\beta_0)}\right) \ = \ \sum_{j=0\
}^2 \sum_{i} \
\frac{\left( \Lambda_i(\beta_j) - \displaystyle \frac{a(\beta_j)}{a(\beta_0)} a(\beta_0) \Lams  \right)^2}
{\delta^2(\Lambda_i)} \ ;
\eeq
where the sum over $j$ covers the sets of coupling estimates for the three $\beta$'s
($\beta_0=3.9$, $\beta_1=4.05$, $\beta_2=4.2$), the index $i$
runs to cover the fitting window of momenta to be contained in a region in which 
the slope $R_0 \sim -90$ was found to be constant. $\Lambda_i(\beta_j)$ is obtained
for any $\beta_j$ by requiring the best-fit to a constant;
$c$ results from the best-fit:
it is the Wilson coefficient of the gluon condensate in \eq{alphahNPt}, where 
the leading logarithm correction is now taken into account, where $ a(\beta_0) q_0=4.5$ 
(this means $q_0 \approx 10$ GeV) was chosen.
The function $\chi^2$ is minimised over the functional space defined by the
four parameters that are explicitly put in arguments for \eq{chi2T}'s l.h.s.:
$a(\beta_0)\Lams$, $c$, $\frac{a(\beta_1)}{a(\beta_0)}$, $\frac{a(\beta_2)}{a(\beta_0)}$.
Thus we obtain all at once $\Lams$ and the gluon condensate, in units of the
lattice spacing for $\beta_0=3.9$, and the ratios of lattice spacings for our three
simulations after the extrapolation to the limit $\mu_q \to 0$ (see tab.~\ref{tables}).
The errors are calculated again by jackknife analysis.
The ratios of lattice spacings can be applied to express the momenta
for all the three sets of coupling estimates plotted in fig.~\ref{fig:C0} (left)
in units of the lattice spacing at $\beta=3.9$. Thus they indeed match each other
and fit pretty well to the analytical prediction with the
best-fit parameters for $\Lams$ and the gluon condensate, in units of $1/a(3.9)$
(see tab.~\ref{tables}), as can be seen in the plot of fig.~\ref{fig:C0}~.                                 
A detailed discussion about systematics can be found in~\cite{BlossierGravina}
indicating that main sources of errors are under control.
Assuming the value $a(3.9) \ = \ 0.0801(14)~\mbox{\rm fm}$~\cite{Baron:2009wt}, we quote our result as
\beq\label{final}
\Lams = \left(330 \pm 23\right) \times \frac{0.0801 \ \mbox{\rm fm}}{a(3.9)} \ \mbox{\rm MeV} \ , \, \,
g^2(q_0^2) \langle A^2 \rangle_{q_0} =  \left(2.4 \pm 0.8\right) \times \left(\frac{0.0801 \
\mbox{\rm fm}}{a(3.9)}\right)^2 \ \mbox{\rm GeV}^2 \ . \nonumber
\eeq

\section{Conclusions and outlooks}

We computed the renormalized strong coupling constant analyzing a variety 
of $N_f=2$ gauge configurations generated in the ETM Collaboration.
We performed an elaborated treatement of the lattice artefacts and 
a precise estimate of the couplings at the infinite cut-off limit.
The coupling
estimates for lattices at different $\beta$'s were seen to match pretty well, as
should happen  if the cut-off limit is properly taken, when plotted in terms of
the renormalization momenta converted  to the same units by applying the
appropriate lattice spacings ratios. These ratios could be either taken  from
independent computations or obtained by requiring the best matching with pretty
compatible results.
Thus, once we are left with the estimates of the coupling constant extrapolated
at vanishing dynamical mass $\mu_q$, for every value   of the renormalization
momentum, $\mu$, they were converted via a fit with a four loops formula into
the  value of $\Lams$. As in th {\it quenched} case a condensate $\langle A^2 \rangle$
is needed in order to get a constant $\Lams$.
As an outlook, we want to apply the same analysis to the case of
lattice QCD with $N_f=2+1+1$ and $N+f=4$ dynamical flavors. 
This will lead to give a reliable lattice prediction for the coupling 
constant, say at $M_Z$, to be compared with available experimental 
determinations.

\section*{Acknowledgements}
We thank the IN2P3 Computing Center
(Lyon) and the apeNEXT computing laboratory (Rome)
where part of our simulations have been done.
J. R-Q is indebted to the Spanish MICINN for the
support by the research project FPA2009-10773 and
to ``Junta de Andalucia'' by P07FQM02962.

\end{document}